\documentclass[twocolumn,aps,preprintnumbers,amsmath,amssymb,floatfix,showkeys]{revtex4}
\usepackage{graphicx}

\begin{document}

\title{Thermomagnetic instabilities in Nb films deposited on glass substrates}
\author{I. Abaloszewa}
\email{abali@ifpan.edu.pl}
\author{M. Z. Cieplak}
\author{A. Abaloszew}
\affiliation{Institute of Physics, Polish Academy of Sciences, 02-668 Warszawa, Al. Lotnik{\'o}w 32/46, Poland}

\date{\today}

\begin{abstract}
In this work, we provide a systematic study of the magnetic field penetration process and avalanche formation in niobium films of different thicknesses deposited on glass substrates. The research was carried out by means of direct visualization of the magnetic flux using magneto-optical imaging. The experimental data were compared with theoretical predictions for the development of thermomagnetic instabilities in the form of finger or dendritic flux avalanches in thin films. Analysis of the temperature and thickness dependence of threshold magnetic field at which superconductor first becomes unstable, as well as the flux penetration depth corresponding to this field allows the evaluation of the thermal and superconducting parameters of the studied films, such as heat transfer coefficient across the film-substrate boundary, thermal conductivity, critical current density.

\end{abstract}\maketitle

\section{Introduction}
Inhomogeneous magnetic field penetration in superconductor films, especially in the form of thermomagnetic instabilities and avalanches, is still an important subject of intense study and analysis. These phenomena can limit the range of applications or even hinder the use of superconducting thin films in electronic devices. In a type-II superconductor, once the external magnetic field reaches a value equal to the first critical field H$_{c1}$, magnetic flux begins to penetrate into the superconductor in the form of quantized flux lines or vortices. As the field increases, the vortices gradually propagate inside the superconductor, causing local heat release due to the motion of the normal vortex core. In some places, due to local thermal and electromagnetic fluctuations, positive feedback can occur: if the system is not able to redistribute the released heat fast enough, overheating will develop in these places, causing local weakening of the pinning force of the vortices and inward propagation of magnetic flux at high speed in the form of narrow dendritic avalanches. Up to now, numerous theoretical and experimental studies of avalanche behavior in films of various superconductors have been carried out \cite{Welling, Aranson, Denisov1, Denisov2, Carmo, Baruch-El, BlancoAlvarez, Pinheiro}.

As shown in an extended theoretical research perfectly matched by an experimental results on MgB$_2$ thin film strips of different widths carried out by Denisov \textit{et al}  \cite{Denisov1, Denisov2}, thin films are more unstable than bulk superconductors and the thermal characteristics of both the superconductor and the substrate, as well as the properties of the interface between them play a crucial role in the development of such instabilities in thin films. In these works it has been shown that the threshold field for the first instability $H_{\rm{th}}$ in a superconducting strip of width $2w$ and thickness $d \ll w$, characterized by critical current density $j_{c}$ and superconducting critical temperature $T_c$, expresses as

\begin{equation}
H_{\rm{th}} = \frac{j_{c}d}{\pi}\arccos\left({\frac{w}{w-l_{\rm{th}}}}\right).  \label{Hth}
\end{equation}
 
Here $l_{\rm{th}}$ is the flux penetration depth corresponding to $H_{\rm{th}}$ field,
 
\begin{equation}
l_{\rm{th}} = \frac{\pi}{2}\sqrt{\frac{\kappa}{\lvert j'_c\rvert E}}\left(1-\sqrt{\frac{2h_0}{nd\lvert j'_c\rvert E}}\right)^{-1},  \label{lth}
\end{equation}

where $\kappa$ is the thermal conductivity of the superconducting film, $j'_{c}$ is the temperature derivative of the critical current density, $h_0$ is the heat transfer coefficient between the film and the substrate, parameter $n$ defines nonlinearity of the current-voltage characteristics in the flux creep regime, $E \propto j^n$, where $E$ is an electric field.

Some aspects of the appearance and behavior of avalanches in niobium films as a function of film thickness have been studied by us \cite{Abalosheva} and others \cite{Colauto, Jing, Brisbois}, but the correlation between such thickness-dependent behavior and the thermal properties of the superconducting material and the film-substrate interface has not yet been studied experimentally. In this paper, we report an investigation of both temperature and film thickness dependence of the $H_{\rm{th}}$ and $l_{\rm{th}}$ in Nb films deposited on the glass substrates, and on the basis of the above theoretical model \cite{Denisov1, Denisov2}, we obtain the thermal conductivity, the film-to-substrate heat transfer coefficient or Kapitza thermal boundary conductance (TBC), and the critical current density of our films.

\section{Sample preparation and measurement details}

Nine niobium films ranging in thickness from 400 nm to 1400 nm was grown by DC magnetron sputtering process at room temperature on glass substrates. The Nb film thickness was controlled by deposition time after the deposition rate has been determined from low-angle x-ray reflectivity measurements. The resulting film set was shaped by photolithography and subsequent reactive ion etching in SF$_6$ plasma into $1.5\times4.5$ mm, $2.5\times2.5$ mm and $1.5\times2$ mm rectangles.

The visualization of magnetic flux penetration into the superconducting films was carried out by magneto-optical imaging (MOI) method. Samples were placed inside a continuous-flow cryostat (temperature in the range of 4 - 300 K) equipped with a low-magnetic field Helmholtz coil. The measurements were performed using an iron-garnet indicator placed directly on the top of the sample, which due to the Faraday effect rotate the plane of polarization of linearly polarized light proportional to the component of the local magnetic field in the direction of propagation, thus allowing to visualize the penetration of magnetic flux into the superconducting film. The image of the magnetic flux distribution was then recorded using a polarization microscope and a CCD camera and transferred to a computer for further processing. 

To determine the temperature dependence of the resistance of the samples, standard four-probe DC transport measurements have been carried out in the temperature range from 4.2 to 300 K.

\section{Results and discussion}

\begin{figure}
\includegraphics[width=8cm]{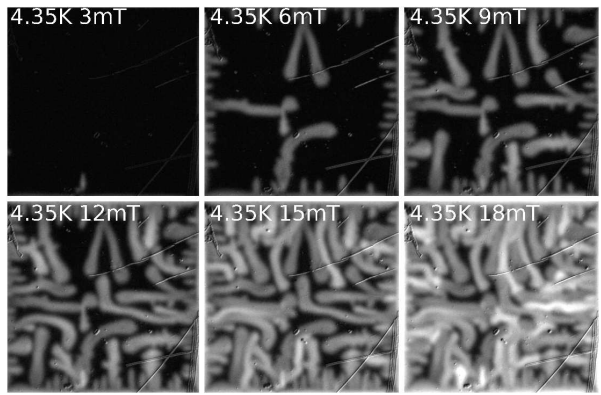}
\caption{MO images of flux penetration in zero-field cooled 1000 nm Nb film shaped in $2.5\times2.5$ mm square.}
\label{Fig1}
\end{figure}

Figure \ref{Fig1} shows magneto-optical images of flux penetration in zero-field cooled Nb film when the perpendicular external magnetic field ramps up. It can be seen that the flux penetrates the film in the form of both finger and dendritic instabilities.

\begin{figure}
\includegraphics[width=8cm]{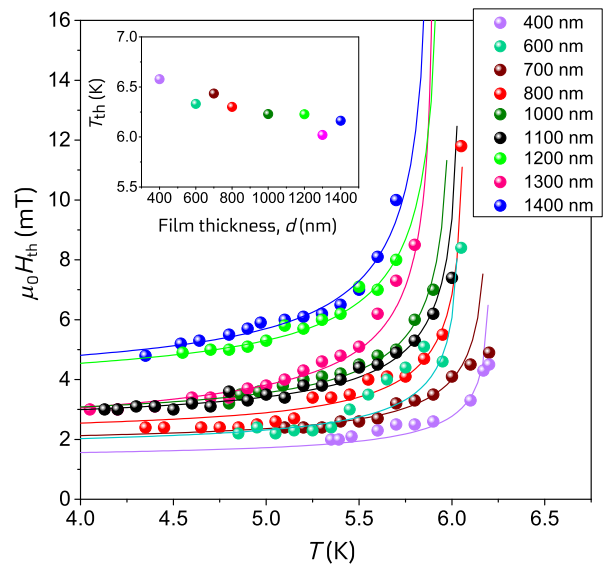}
\caption{Temperature dependences of the threshold magnetic field for the first instability in Nb films of different thicknesses (symbols). The measurements were made on $1.5\times4.5$ mm rectangle Nb films of nine thicknesses. The full lines are theoretical fits according to Eq.~\ref{Hth}, the line color corresponds to the color of the symbols representing the experimental data. The inset shows the threshold temperature above which magnetic flux penetrates the film with no instabilities.}
\label{Fig2}
\end{figure}

The dependences, as functions of the temperature, of the threshold field for the first instability $H_{\rm{th}}$, for nine film thicknesses are shown in Fig. \ref{Fig2}. It is clearly visible that $H_{\rm{th}}$ depends on the sample thickness and there is also a threshold temperature $T_{\rm{th}}$ above which the flux penetrates the film without forming avalanches.  The theoretical fits denoted in Fig. \ref{Fig2} by full lines were obtained by combining Eqs. \ref{Hth} and \ref{lth} and using $T_c = 9.2$ K, $w = 0.75$ mm, $E = 200$ mV/m \cite{Denisov2} and the following assumptions for the temperature dependences of the model parameters $j_c$, $\kappa$, $h_0$ and $n$.  The temperature dependence of the critical current density is taken to be linear, $j_c=j_{c0}(1-T/T_c)$. 
The thermal conductivity of niobium at temperatures below the $T_c$ is described by the theory developed for metals, with certain modifications introduced by the superconducting state: part of the electrons at temperatures below the critical temperature form Cooper pairs, which are not involved in heat transfer, at temperatures below 1.8 K the phonon contribution to the thermal conductivity becomes dominant. Within the temperature range of our experiment ($4 - 9$ K), both electronic and lattice thermal conductivity terms have to be taken into account and $\kappa(T)$ can be approximated by a third order function  $\kappa=\tilde{\kappa}(T/T_c)^3$ \cite{Koechlin}. According to the acoustic mismatch model and diffuse mismatch model \cite{Swartz}, the TBC also obeys the $T^3$-dependence, confirmed experimentally for many interfaces:  $h_0=\tilde{h}_0(T/T_c)^3$. Finally, the current-voltage dependence exponent $n \sim U/kT$ takes the form of $\tilde{n}(T_c/T-1)$ at the pinning potential for vortices $U \propto 1-T/T_c$. We, following \cite{Denisov2} assume $\tilde{n}=40$.
            
\begin{figure}
\includegraphics[width=8cm]{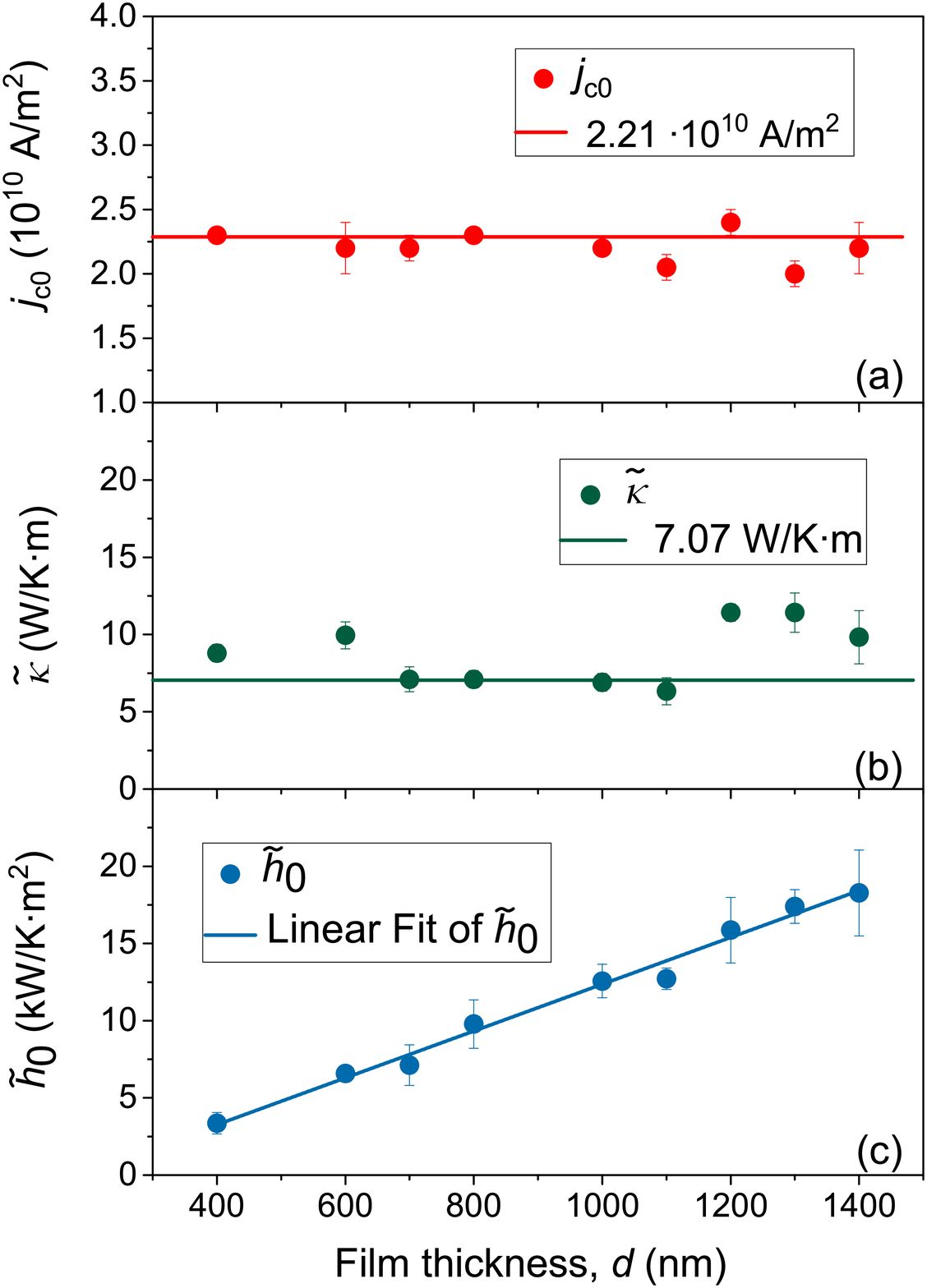}
\caption{Thickness dependences of fitting values of the $j_{c0}$ (the critical current density at $T = 0$) (a), the values of the parameters $\tilde{\kappa}$ (left axis) (b) and $\tilde{h}_0$ (c) (the thermal conductivity  and the heat transfer coefficient at the temperature equal to $T_c$, respectively).}
\label{Fig3}
\end{figure}

The parameters $j_{c0}$, $\tilde{\kappa}$, $\tilde{h}_0$ in the above relations are treated as fitting parameters, which we extract from the fits of the theoretical curves to experimantal data. Figure \ref{Fig3} shows values of the critical current density at $T = 0$ and the values of the thermal conductivity and the heat transfer coefficient at the temperature equal to the $T_c$ as functions of sample thickness.
The $j_{c0}$ is found to be thickness independent and its mean value (of about $2.2\times10^{10} \rm{A/m}^2$, denoted by the full line in the Fig. \ref{Fig3}(a)) is typical for niobium thin films.

For comparison, we can estimate the $j_{c0}$ on the basis of the value of the first penetration field $H_{\rm{p}}$ and the magnetic field penetration depth $\lambda$  at $T = 0$. We determine the lower critical field $H_{\rm{c1}}$  by examining the $H_{\rm{p}}$ similar to the procedure described in \cite{Okazaki}, using the magneto-optical determination of the field at which the remanence at the sample edges first occurs. Since our sample is a thin film, we use Brandt's approximation of $H _{\rm{c1}}$ \cite{Brandt} to include the demagnetizing effect:

\begin{equation}
H_{\rm{c1}} = H_{\rm{p}}/\tanh\sqrt{0.67\times 2w/d}
\end{equation}

The experimental points are fitted by the formula $H_{\rm{c1}} = H_{\rm{c1}}(0)[1-(T/{T_c})^2]$ describing the temperature dependence of  $H_{\rm{c1}}$ in the Meissner state of the sample. This allows us to calculate $\mu _0 H_{\rm{p}}(0)$ and $\mu _0 H_{\rm{c1}}(0)$, which are equal to 1.417 mT and 109.4 mT, respectively.

The London penetration depth $\lambda$ is estimated using the London formula for $\lambda/\xi\approx1$ case \cite{Hu}

\begin{equation}
\mu _0 H_{\rm{c1}} ={\frac {\Phi_0} {4\pi \lambda^2}}\ln {\frac{\lambda}{\xi}},
\end{equation}

where $\Phi_0$ is the flux quantum and $\xi$ is the coherence length, which for niobium is equal to 38 nm. From this we find for that $\lambda(0)$ is equal to 45.3 nm. Finally, from $j_{c0}=H_{\rm{p}}(0)/\lambda(0)$ \cite{London} we obtain that $j_{c0}$ is about $2.43\times10^{10} \rm{A/m}^2$, which is close to the value obtained from fitting $H_{\rm{th}}$.

The $\tilde{\kappa}$ parameter does not show a pronounced thickness dependence either (Fig. \ref{Fig3}(b)). As mentioned above, in the superconducting state, paired electrons in niobium decouple from the lattice and no longer participate in heat conduction. Heat is then carried by the thermally-excited quasiparticles and phonons. The poor thermal conductivity of niobium is thus intrinsically due to its superconducting nature. The thermal conductivity values extracted for our niobium films are quite low compared to those obtained on high quality films \cite{Koechlin}, indicating a large number of defects arising during sample deposition. We may compare this low value of thermal conductivity with the value extracted form residual resistivity ratio, $RRR=\rho_{273\rm{K}}/\rho_{4.2\rm{K}}$, where $\rho_{4.2\rm{K}}$ is determined by extrapolation of $\rho(T)$ curve to $4.2$~K. Using $RRR$ of 1000 nm thick Nb film and taking into account that the thermal conductivity is approximately $RRR/4$ at $4.2$~K \cite{Kadanoff, Padamsee}, we obtain: $RRR=\rho_{273\rm{K}}/\rho_{4.2\rm{K}} = 2.69 = 4 {\kappa}\rvert_{T=4.2\rm{K}}= 4 \tilde{\kappa}\times(4.2~\rm{K}/ 9.2~\rm{K})^3\Rightarrow \tilde {\kappa} =7.07~\rm{W}/(\rm{K} \times \rm{m}).$ This value, denoted by the full line in the Fig. \ref{Fig3}(b), is in excellent agreement with the lower range of the extracted $\tilde{\kappa}$ values.

Finally, we find that the film-to-substrate heat transfer coefficient $\tilde{h}_0$ grows linearly with thickness and is well described by a straight line (Fig. \ref{Fig3}(c)): $\tilde{h}_0=ad-b$, with $a=1.52\times10^{7}$ and $b=2.8$. Trying to understand this result we note that we can treat our films as a two-part solid-solid system, as has been shown in the work \cite{Brisbois} that the metal mirror of the indicator placed on the sample during magneto-optical measurements does not influence the heat distribution in the underlying layers if the distance between the indicator and film surfaces is on the order of one micron. This condition is fulfilled in our experiments due to the presence of microscopic residual photoresist used for photolithographic sample shaping.

A similar Kapitza TBC behavior, i.e. an increase in heat transfer across film-substrate boundary with film thickness has been observed in experiments with other solid-solid interfaces. For example, in the work \cite{Yuan} TBC increase between MoS$_2$ and Si with thickness is explained by improved mechanical stiffness of thicker samples and the resulting better interface contact. Our Nb films were deposited at room temperature and as the deposition time increased to obtain thicker samples the temperature of the films did not increase enough for an annealing effect to occur at the interface. Such a change could have occurred if the films had been heated above $100^\circ$C, but during the deposition of our samples the film was in thermodynamic equilibrium with the substrate and water-cooled holder. Thus, we assume that the interfacial conditions are approximately the same for samples of different thicknesses.

In works \cite{Kuo, Schafft} it has been reported that the thinner Si and SiO$_2$ films deposited on single-crystalline Si substrate show lower overall (from thin film and interface) thermal conductivity value than thicker ones. This is explained by the influence of phonon scattering at the interface (film thickness-dependent), on the thermal conductivity of the films themselves rather than on the heat transfer coefficient to the substrate. We are dealing with a somewhat different situation, since in niobium, in the temperature range in which we carried out the measurements, the thermal conductivity is determined by both electron and lattice contributions and, as shown above, is independent of the thickness.

In other works \cite{Ong, Majee} it has been found that the Kapitza resistance of the graphene-SiO$_2$, MoS$_2$-SiO$_2$, and WSe$_2$-SiO$_2$ interface decreases (or the TBC increases) with the thickness of the film. This is a result of the relaxation of tensile strain which occurs with increasing number of monolayers of the film-forming material on SiO$_2$, which leads to more efficient transmission of the higher-frequency phonons from multilayer structure to substrate. Our films are rather thick and were deposited on an amorphous substrate, so no strain relaxation effects occur. Therefore, we need to look for other possible explanation of our result.

In general, TBC in such systems is determined by the scattering of thermal energy carriers, i.e. electrons and phonons, at the interface between two solids \cite{Swartz}. At low temperatures, the heat transfer coefficient through the niobium-glass interface is mainly determined by phonon scattering at the film-substrate interface and phonon transfer across this interface, as the effect of electrons on the TBC is negligible \cite{Anderson, Majumdar}. In the acoustic mismatch model (AMM) \cite{Swartz} the interface is perfectly smooth and phenomena such as Snell's acoustic laws, critical angle and the total internal reflection are inherent to the behavior of phonons in materials on either side of the interface. Since the phonon velocity in niobium is lower than in glass due to the higher stiffness and lower density of glass, there is a critical angle of incidence for phonons in niobium above which phonons experience total internal reflection and do not pass into the substrate. The probability of phonon reflection from the boundary and passing through it depends, in turn, on the difference in acoustic impedance of the materials, $z=\rho c$, where $\rho$ is the mass density and $c$ is the sound velocity in the material.

According to diffuse mismatch theory (DMM), while some of the phonons are reflected specularly from the film-substrate boundary, others are diffusely scattered on the boundary irregularities. An increase in the number of phonons incident on the boundary at a small angle with respect to the normal after re-reflection from the surfaces leads to an increased probability of phonon transmission into the substrate, and hence a reduction in the Kapitza resistance and an increase in the TBC between the materials \cite{Schmidt, Yang}. Reduction of the sample thickness limits the propagation of long-wave phonons and the entrance of these phonons into the acceptance cone, which causes the film to act as a filter for phonon spectra.

\begin{figure}
\includegraphics[width=8cm]{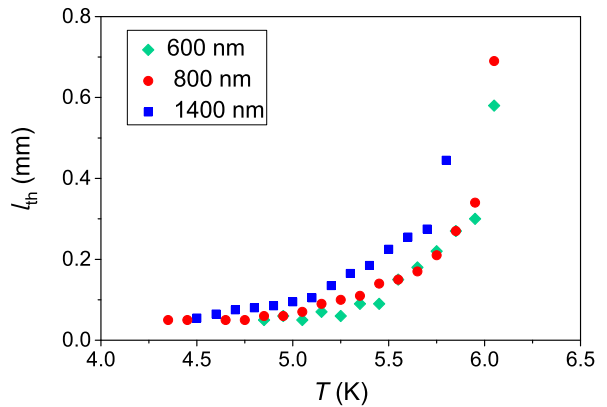}
\caption{Temperature dependences of the threshold depth $l_{\rm{th}}$ for three Nb films of different thicknesses.}
\label{Fig4}
\end{figure}

The thickness dependence of the $\tilde{h}_0$ is confirmed by the fact that there is almost no such dependence of $l_{\rm{th}}$, as can be seen from the Eq.~\ref{lth} and experiment. Figure \ref{Fig4} shows that experimental temperature dependences of the $l_{\rm{th}}$ for three Nb films of different thicknesses are very similar, except in the area $T\rightarrow T_{\rm{th}}$, where the $l_{\rm{th}}$ is discontinuous. This can be understood from a mathematical point of view if we accept that $\tilde{h}_0(d)$ includes constant $b$. The relatively small negative value of $b$ reduces the value of $\tilde{h}_0$ by  2.8 $\rm{kW}/\rm{K} \times \rm{m}^2$. It is possible that its presence is caused by inelastic scattering of a part of phonons on imperfections at the interface (such as roughness or atomic mixing), which does not depend on the size of the sample. The presence of $b$ has relatively minor influence on the behavior of the slow-growing part of $l_{\rm{th}}(T)$ for different thicknesses, but begins to affect the fast-growing part of the dependence significantly.

To estimate the $T_{\rm{th}}$ values, it is necessary to find the point at which $H_{\rm{th}}$ has a discontinuity.  If the $l_{\rm{th}}$ was not limited by the size of the sample, it would be discontinuous at a point satisfying the condition $2h_0/(nd\lvert j'_c\rvert E)=1.$ Substituting the temperature dependencies of the parameters comprising the above expression, we obtain the equation $T^4+\alpha(T-T_c)=0,$  where $\alpha=\tilde{n}{T_c}^{2} d j_{c0} E/(2\tilde{h}_0).$

$H_{\rm{th}}$ undergoes a discontinuity at the point where the fast-growing function $l_{\rm{th}}$, approaching its discontinuity, being a solution of  the equation above, reaches a value of $w$. Thus, the sample width reduction leads to a decrease in $T_{\rm{th}}$ compared to the situation when $w\rightarrow\infty$. The inset in Fig.\ref{Fig2} shows the numerically extracted values of $T_{\rm{th}}$. The calculation was done with the parameters used in the theoretical fits of the $H_{\rm{th}}$. In the case of absence of the term $b$ (i.e. direct proportionality $\tilde{h}_0=a d$), the temperature $T_{\rm{th}}$ would be constant and would be about 6.01~K.

\section{Conclusions}

We have used a non-invasive and non-destructive magnetooptical method to determine the thickness dependence of parameters characterizing the thermal and superconducting properties of niobium thin films in which thermomagnetic instabilities develop. Based on a comparison of the theoretical model describing the behavior of the threshold field of the first jump with our experimental data, we determined that the thermal conductivity and critical current density of our films are thickness-independent. On the other hand, the heat transfer from the superconducting film to the substrate depends linearly on the film thickness. The linear behavior can be explained in the frame of the diffuse mismatch model. We have found that the field penetration depth into the sample, at which the first flux avalanche occurs, is practically independent of the specific dimensions of the sample (except that it is limited to half the width of the sample). However, it is determined by the material properties that compose the film-substrate system. Understanding the mechanisms and ways of changing heat transfer can be helpful in the goal of controlling TBC and thermomagnetic instabilities appearance in engineering applications of superconducting films.

\section*{Acknowledgements}
We are grateful to L. Y. Zhu and C.-L. Chien (Johns Hopkins University) for growing the films used in this study.

\end{document}